\documentclass[10pt,aps,prb,raggedbottom,longbibliography,reprint,citeautoscript,letterpaper,superscriptaddress]{revtex4-2} 
\usepackage[usenames,dvipsnames]{color}
\usepackage{graphicx}
\usepackage[bookmarks=false,colorlinks]{hyperref}
\hypersetup{linkcolor=magenta,citecolor=MidnightBlue,filecolor=Plum,urlcolor=MidnightBlue}
\usepackage[all]{hypcap} 
\usepackage{lmodern}



\begin{document}

\title{Canted Antiferromagnetism in Polar MnSiN$_2$ with High N\'eel Temperature}

\author{Linus Kautzsch}
\email{kautzsch@ucsb.edu}
\affiliation{Materials Department and Materials Research Laboratory,
University of California, Santa Barbara, Santa Barbara, CA, 93106}

\author{Alexandru B.\ Georgescu}
\email{georgesc@iu.edu}
\affiliation{Department of Materials Science and Engineering, Northwestern University, Evanston, Illinois  60208, USA}
\affiliation{Department of Chemistry, 800 East Kirkwood Avenue, Indiana University, Bloomington, Indiana 47405, United States}

\author{Danilo Puggioni}
\affiliation{Department of Materials Science and Engineering, Northwestern University, Evanston, Illinois  60208, USA}

\author{Greggory Kent}
\affiliation{Materials Department and Materials Research Laboratory,
University of California, Santa Barbara, Santa Barbara, CA, 93106}

\author{Keith M. Taddei}
\affiliation{Neutron Scattering Division, Oak Ridge National Laboratory, Oak Ridge, TN, 37831}

\author{Aiden Reilly}
\affiliation{Materials Department and Materials Research Laboratory,
University of California, Santa Barbara, Santa Barbara, CA, 93106}

\author{Ram Seshadri}
\affiliation{Materials Department and Materials Research Laboratory,
University of California, Santa Barbara, Santa Barbara, CA, 93106}
\affiliation{Department of Chemistry and Biochemistry, 
University of California, Santa Barbara, Santa Barbara, CA, 93106}

\author{James M.\ Rondinelli}
\email{jrondinelli@northwestern.edu}
\affiliation{Department of Materials Science and Engineering, Northwestern University, Evanston, Illinois  60208, USA}

\author{Stephen D.\ Wilson}	
\email{stephendwilson@ucsb.edu}
\affiliation{Materials Department and Materials Research Laboratory,
University of California, Santa Barbara, Santa Barbara, CA, 93106}

\begin{abstract}
MnSiN$_2$ is a transition metal nitride with Mn and Si ions displaying an ordered distribution on the cation sites of a distorted wurtzite-derived structure. The Mn$^{2+}$ ions reside on a 3D diamond-like covalent network with strong superexchange pathways. We simulate its electronic structure and find that the N anions in MnSiN$_2$ act as $\sigma$- and $\pi$-donors,  which serve to enhance the N-mediated superexchange, leading to the high N\'{e}el  ordering temperature of $T_N$ = 443\,K. Polycrystalline samples of MnSiN$_2$ were prepared to reexamine the magnetic structure and resolve previously reported discrepancies. An additional magnetic canting transition is observed at $T_\mathrm{cant}$ = 433\,K and the precise canted ground state magnetic structure has been resolved using a combination of DFT calculations and powder neutron diffraction. The calculations favor a $G$-type antiferromagnetic spin order with  lowering to $Pc^\prime$. Irreducible representation analysis of the magnetic Bragg peaks supports the lowering of the magnetic symmetry. The computed model includes a 10$^\circ$ rotation of the magnetic spins away from the crystallographic $c$-axis consistent with measured powder neutron diffraction data modeling and a small canting of 0.6\,$^\circ$. 
\end{abstract}

\date{\today}
\maketitle

\section{Introduction}
Ternary nitrides are a burgeoning materials class hosting diverse structures, compositions, and properties that make them appealing for various applications \cite{greenaway21}. 
Most of this research focuses on semiconducting and metallic nitrides for optoelectronic, piezoelectric, lighting, and structural applications, harnessing their cation-tunable electronic structures and superior mechanical properties \cite{Sun19,Ashraf20,Brinkley11,Jena19,Zerr06}. 
However, discovery of novel ternary nitrides in new compositions is challenging owing to difficulty in breaking the strong molecular dinitrogen bonds and overcoming thermodynamic driving forces that favor simpler binary nitrides.
Nonetheless, recent successful syntheses have led to the discovery of bulk CaSnN$_2$ and CaTiN$_2$ \cite{Kawamura21,Li17}, the thin films MgTiN$_2$, ZnMoN$_2$, and ZnZrN$_2$ \cite{Sun19,Arca18,Woods22} and  MgSnN$_2$ and MgZrN2$_2$ in both thin-films and bulk samples \cite{Greenaway20,Kawamura20,Bauers19,Rom21}.

Magnetic nitride-rich transition metal compounds generally remain underexplored, despite the unique chemical bonding afforded by the N$^{3-}$ ion with open $d$-shell transition metals.
The nitride anion has a higher oxidation state and lower electronegativity than the oxide O$^{2-}$ anion and prefers 3-coordinate, triangular bonding preferences\cite{vennos1990synthesis,vennos1992synthesis,cordier1990ca6gan5,kloss2021preparation} rather than 2-coordinate linear bonding preferences of the oxide anion.
The local bond geometries are further stabilized by stronger $\pi$-bonding interactions, which because of the lower electronegativity of nitrogen also result in shorter metal-nitride bond lengths.
This increases the covalency of the magnetic cation $-$ nitrogen ligand bond and can lead to low-spin\cite{vennos1990synthesis} or non-magnetic configurations\cite{vennos1992synthesis}. However, in some cases this increased covalency favors higher magnetic ordering temperatures.\cite{wintenberger1972etude,Wintenberger77,coey1999magnetic}
The resulting magnetic coupling, paired with nitrogen's preference for 3-coordinate bonding, may lead to novel magnetic phenomena in low-dimensional crystal structures forming triangular and hexagonal networks. 
Such phenomena include noncollinear spin structures\cite{hajiri2021spin} and symmetry-broken ground states or longer range magnetic interactions which may promote frustration and suppress magnetic order.\cite{wang2020ferromagnetic,trocoli2022mnta2n4}

Antiferromagnetic (AFM) semiconductors with ordering temperatures above room temperature are of increased technological interest owing to advances in AFM spintronics \cite{Jungwirth_2016},  current-induced switching of AFM \cite{Wadley_2016}, noncollinear spin-splitting \cite{PhysRevMaterials.5.014409,PhysRevB.101.220403}, and magnetoelectric memories \cite{Kosub_2017,Spaldin_2019}.
Promising members of this class of materials are manganese nitrides, which form a family exhibiting oxidation states ranging from Mn$^{5+}$ ($d^2$ electronic configuration) to Mn$^{2+}$ ($d^5$) and can host large magnetic moments due to their large Hund's coupling on the Mn $d$ manifold that favors high-spin states \cite{Niewa02}.
The manganese nitridosilicate MnSiN$_2$ adopts the anti-$\beta$-NaFeO$_2$-type structure, distorted from wurtzite, with ordering of divalent Mn and tetravalent Si cations.
The chemical ordering produces a lattice orthorhombicity \cite{Breternitz_2021} within the polar noncentrosymmetric space group $Pna2_1$ (number 33) and three-dimensional (3D) corner connectivity of MnN$_4$ and SiN$_4$ tetrahedra with the crystal chemical formula ${}^{~3}_{\infty}[\mathrm{MnN}_{4/4}]^-[\mathrm{SiN}_{4/4}]^+$ (\autoref{fig1:diff}a).

Semiconducting MnSiN$_2$ exhibits a high magnetic ordering temperature arising from its  Mn$^{2+}$ ions residing on a 3D diamond-like network. 
It was first synthesized in an amonolysis process \cite{Maunaye71} ; an antiferromagnetic ordering temperature of $T_\mathrm{N}=453$~K was later reported in  \cite{Wintenberger77}. 
Further characterizing the state, Esmaeilzadeh et al.\ employed neutron diffraction and observed a broad magnetic transition \cite{Esmaeilzadeh06}, which they  conjectured arose from frustration of the collinear spin structure aligned along the $c$ axis and low-dimensional character in the system.
This behavior was suggested to be alleviated by disordered spin canting of an unknown amount below $T_\mathrm{N}=453$\,K \cite{Wintenberger77}, and a Curie-Weiss constant much greater than $T_\mathrm{N}$ was reported indicating strong magnetic fluctuations above $T_\mathrm{N}$.
Until now, all models of the magnetic structure have only explored those magnetic configurations compatible with the nuclear symmetry of MnSiN$_2$ \cite{Wintenberger77}.

Here, we reexamine the magnetic structure of MnSiN$_2$ and resolve the magnetic discrepancies reported previously. 
We perform density functional theory (DFT)  calculations to guide our magnetic structure refinement of temperature-dependent neutron diffraction data and also show that the high N\'eel temperatures arise from  strong Mn-N $\pi$-bonding interactions. 
We find $T_\mathrm{N}=443$\, K in our samples and this initial state corresponds to AFM collinear spin ordering along the crystallographic $c$ axis with an orthorhombic magnetic symmetry, $Pn^\prime a2_1^\prime$.
Upon cooling, we then find magnetic canting at $T_{\rm cant}=433$\,K which arises from a magnetic symmetry breaking to the monoclinic $Pc^\prime$ symmetry, such that there are two inequivalent Mn sites with long-range AFM ordering of  spins in the $ac$-plane.
Although both polar phases permit weak ferromagnetism (FM) through a relativistic Dzyaloshinskii-Moriya interaction, only the noncollinear spin state with canted moments in the $Pc$ structure produce weak-FM.
Owing to unintentional magnetic oxide impurities in our samples, we are unable to experimentally resolve the weak-FM. 
Last, we show that the separation in temperature between $T_\mathrm{N}$ and $T_{\rm cant}$ is not a consequence of a structural symmetry lowering but rather arises from entropic competition among competing noncollinear states from $d^5$ Mn in a distorted tetrahedral environment. 

\section{Methods}

\subsection{Sample preparation}

Polycrystalline samples of MnSiN$_2$ were prepared from Mn (Alfa Aesar) 99.95~\% and Si$_3$N$_4$ (Fisher Scientific) 98.5~\% in a nitrogen gas flow reaction \cite{Esmaeilzadeh06}.
The starting materials were mixed using a Mn:Si ratio of 1.1:1, intimately ground, pelletized and heated under a nitrogen gas flow (flow rate $\approx$~10~dm$^3$~min$^{-1}$) for 24\,h at $T=1250$~$^{\circ}$C. 
The samples were then re-ground and pelletized and heated under the same condition for 24 h. Samples with a total mass of 5\,g were prepared. This procedure resulted in well-crystallized samples.

\subsection{Characterization}

The sample quality and purity after each of the two heating steps was verified by powder X-ray  diffraction (XRD) using a Panalytical Empyrean powder diffractometer operating with Cu-K$_{\alpha}$ radiation operating in Bragg-Brentano geometry. High-resolution synchrotron powder X-ray diffraction was collected at the beamline 11-BM at the Advanced Photon Source (APS), Argonne National Laboratory, using an average wavelength of $\lambda=0.458118$\,\AA\ at $T=295$\,K.

Rietveld refinement of the crystal structure was carried out using TOPAS academic. A Thompson-Cox-Hastings pseudo-Voigt profile function (six parameters) was used. 
The Finger model was used to handle peak asymmetry due to the axial divergence of the beam \cite{Finger1994}. 
Unit cell parameters, atomic positions, and isotropic displacement parameters ($B_{\rm iso}$) of the Mn and Si atoms were refined.

\begin{figure}
    \centering
    \includegraphics[width=0.5\textwidth,trim=0 20 0 0, clip]{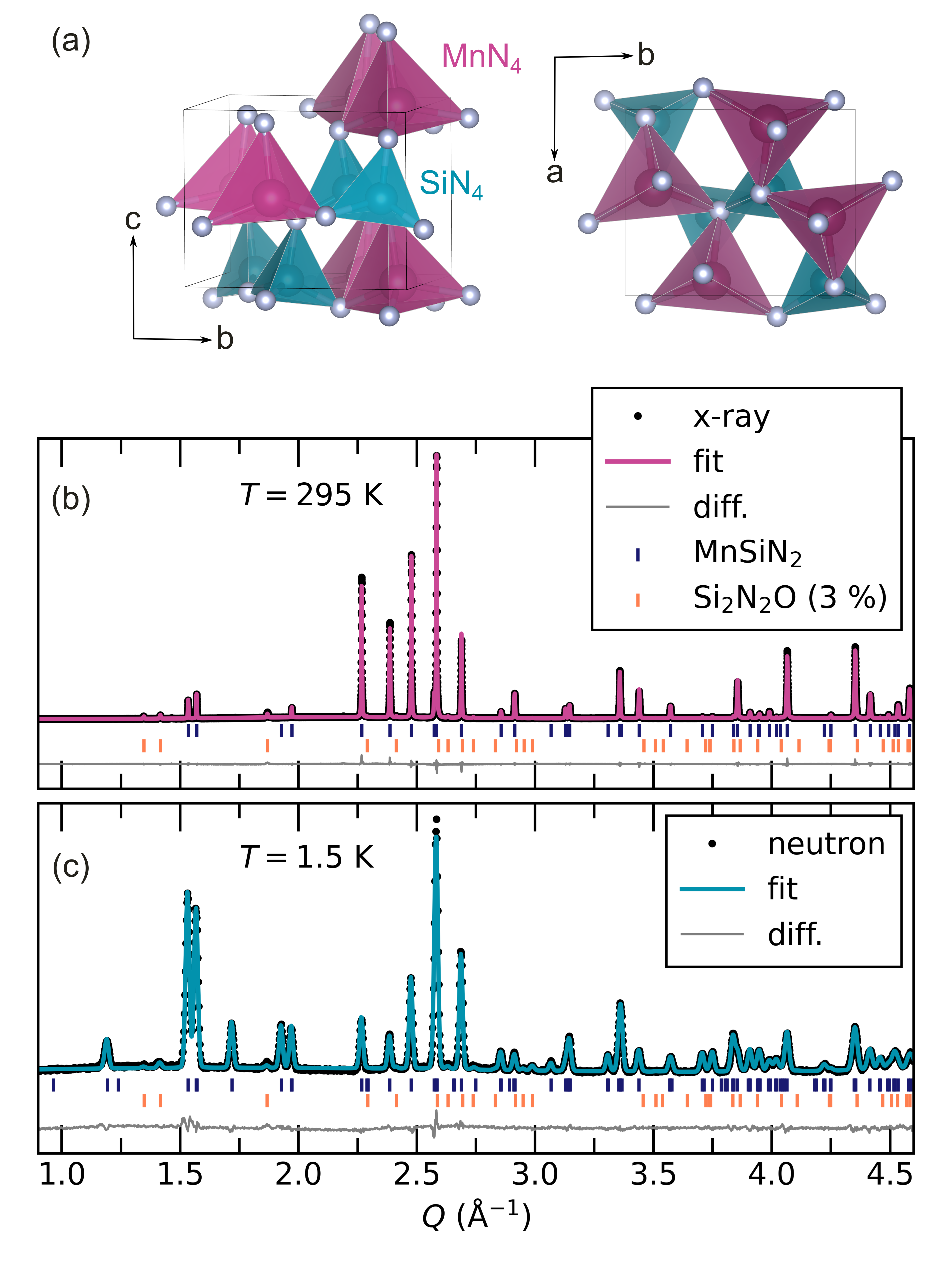}
    \caption{
    (a) Crystal structure of the polar orthorhombic nitride MnSiN$_2$ (space group $Pna2_1$), showing a network of corner-sharing MnN$_4$ and SiN$_4$ tetrahedra.
    (b) Synchrotron x-ray diffraction at $T=295$~K and Rietveld fit to the data using a model consisting of MnSiN$_2$ and the 3~\% impurity phase Si$_2$N$_2$O. 
    (c) Powder neutron diffraction at $T=1.5$~K and Rietveld fit including the crystallographic phases MnSiN$_2$ and Si$_2$N$_2$O and the DFT-calculated magnetic $Pc^\prime$ structure.}
	\label{fig1:diff}
    \end{figure}

Variable temperature powder XRD measurements were collected on a Bruker KAPPA APEX II diffractometer equipped with an APEX II CCD detector using a TRIUMPH monochromater with a Mo K$_{\alpha}$ X-ray source ($\lambda = 0.71073$\,\AA). Finely ground MnSiN$_2$ was sealed in a Kapton tube (0.75~mm diameter) with epoxy and the sample temperature was controlled using an Oxford nitrogen gas cryostream system.  
Each powder XRD pattern was collected using two phi scans set at $-15^{\circ}$ and $-30^{\circ}$ $2 \theta$ (Mo K$_{\alpha}$), with a detector distance of 100\,mm, and utilizing a $t=60$\,s scan rate. Scans were collected between $T=420$\,K\ and $T=462$\,K in 3\,K steps, and subsequently merged and integrated using the APEXII software suite.

Powder neutron diffraction data were collected on the HB-2A beamline at the High Flux Isotope Reactor (HFIR), Oak Ridge National Laboratory. 
A Ge(113) monochromator and a wavelength of $\lambda=2.4067$\,\AA\ was used. Diffraction data were collected at temperatures of 1.5~K, 5~K, 20~K, 30\,K, and 60\,K. 
Rietveld refinements of the neutron diffraction data were carried out using using the Fullprof program \cite{RODRIGUEZCARVAJAL199355}. 
The software package SARAh was used to assess the propagation vector and the irreducible group representations from the observed magnetic scattering \cite{Wills2000}. 

DC magnetic measurements between 2 K and 400 K were performed using a Quantum Design MPMS3 SQUID vibrating sample magnetometer.
Samples with masses of $\approx$~5~mg were measured in polypropylene capsules. Data at temperatures above 400~K were collected with the oven-stick option and samples of $\approx10$\,mg were cemented to the oven-stick. DC magnetization measurements were collected under a constant magnetic field while sweeping temperature at a rate of 7\,K\,min$^{-1}$.

\subsection{Density functional theory}
We performed DFT calculations using the Vienna Ab Initio Simulation Package (VASP) \cite{kresseInitioMolecularDynamics1993,kresseInitioMoleculardynamicsSimulation1994,kresseEfficiencyAbinitioTotal1996,kresseEfficientIterativeSchemes1996} using the recommended projector augmented wave (PAW) pseudopotentials, \cite{kresseUltrasoftPseudopotentialsProjector1999,blochlProjectorAugmentedwaveMethod1994} and the GGA-PBE exchange correlation functional \cite{perdewGeneralizedGradientApproximation1997} with the following valence configurations: 
Mn ($3p^64s^23d^5$), 
Si ($3s^23p^2$), and
N ($2s^2p^3$).
We used a 520\,eV energy cutoff for the plane-wave expansion and an energy convergence of $10^{-8}$\,eV/16 atom formula unit (f.u.) energy difference between consecutive self-consistent steps.
For $k$-point integrations, we used Gaussian smearing ($\sigma=0.05$\,eV), and sampled the Brillouin zone with a $9\times8\times7$ $k$-point mesh. 
The atomic positions were relaxed until forces were less than 0.05\,meV/\AA.

Multiple collinear and noncollinear spin configurations were explored by including spin-orbit interactions.
We performed a sensitivity analysis of the potential magnetic ground states and band gap  
with alternative exchange-correlation functionals, including the strongly constrained and
appropriately normed (SCAN) meta-GGA \cite{PhysRevLett.115.036402} and PBE plus a Hubbard $U$ correction ($\textrm{PBE}+U$ using both the 
Dudarev \cite{PhysRevB.57.1505} and Anismov \cite{PhysRevB.52.R5467}.
Minor quantitative changes were found; in all cases, MnSiN$_2$ remains an AFM semiconductor, indicating the insulating gap arises from the strong Hund's coupling for the half-filled Mn$^{2+}$ $d^5$ manifold. 
For clarity, we focus on results obtained using $U(\textrm{Mn})=4$\, eV and $J(\textrm{Mn})=1.5$\,eV within the Anisimov $\textrm{PBE}+U+J$ approximation, which provides quantitative agreement with the experimental optical band gap.

\begingroup
\squeezetable
\begin{table}[t]
\centering
\caption{Results of Rietveld refinement of scattering data. Lattice parameters $a$, $b$, and $c$; magnetic moments $\mu_{\rm Mn}$ on the Mn site; $R$-factors $R_{\rm wp}$ and $R_{\rm exp}$; carried out based on high temperature $Pna2_1$ and DFT-calculated $P_c^\prime$ structure. The magnetic structure for the  $Pna2_1$ crystallographic group comprises a   combination of basis vectors from two Shubnikov groups: $(Pna^\prime2_1^\prime,33.147)$ and $(Pn^\prime a2_1^\prime,33.146)$.}
\begin{ruledtabular}
\begin{tabular}{cccccccc}
 &  \multicolumn{1}{c}{X-ray} & \multicolumn{6}{c}{neutron} \\
 \cline{2-2}\cline{3-8}
{\textit{T} (K)} & 295 & \multicolumn{2}{c}{60} & \multicolumn{2}{c}{20} & \multicolumn{2}{c}{1.5}  \\
\cline{2-2}\cline{3-4}\cline{5-6}\cline{7-8}
{symmetry} & $Pna2_1$ & $Pna2_1$ & $Pc^\prime$ & $Pna2_1$ & $Pc^\prime$ & $Pna2_1$ & $Pc^\prime$ \\
\hline
\textbf{\textit{a} (\AA)} & 5.266 & 5.268 & 5.075 & 5.268 & 5.075 & 5.268 & 5.075\\
\textbf{\textit{b} (\AA)} & 6.519 & 6.518 & 5.268 & 6.518 & 5.268 & 6.518 & 5.268\\
\textbf{\textit{c} (\AA)} & 5.073 & 5.075 & 8.257 & 5.075 & 8.253 & 5.075 & 8.253\\
\textbf{$\beta$ ($^{\circ}$)} & 90 & 90 & 127.86 & 90 & 127.82 & 90 & 127.82 \\
\textbf{$\mu_{\mathbf{Mn}}$ ($\mu_{\mathbf{B}}$)} & $-$ & $-$ & 4.37  & $-$ &  4.37  & $-$ & 4.37\\
\textbf{\textit{R}$_{\mathbf{wp}}$} & 6.28 & 13.6 & 13.4 & 14.2 & 13.7 & 14.6 & 14.5\\
\textbf{\textit{R}$_{\mathbf{exp}}$} & 4.41 & 6.94 & 7.02 & 6.93 & 5.82 & 6.94 & 6.96\\
\end{tabular}
\label{tab:refin}
\end{ruledtabular}
\end{table}
\endgroup

\section{Results and Discussion}

\subsection{Crystal structure and chemical bonding}

\autoref{fig1:diff}b shows our Rietveld refinement of the synchrotron X-ray diffraction data collected at $T=295$~K and confirms MnSiN$_2$ crystallizes in the previously reported orthorhombic 
$Pna2_1$ (\autoref{fig1:diff}a) space group 
\cite{Esmaeilzadeh06}.
We find sharp Bragg peaks indicating excellent crystallinity of the sample. 
A small (3~\% mass) impurity of Si$_2$N$_2$O was included in the fit and the fitting parameters are presented in \autoref{tab:refin}. 
The lattice parameters obtained from Rietveld refinement of the X-ray data and the neutron data are in excellent agreement across the two techniques for space group $Pna2_1$ at all temperatures, indicating negligible thermal contraction from $T=295$~K to $T=1.5$~K.

The experimentally determined 295\,K structure is in close agreement with the DFT-relaxed structure, which has a 1.65\,\% larger unit cell volume. A deviation of the lattice parameters of this magnitude is typical for calculations using the GGA-PBE functional.\cite{zhang2018performance}
The local bonding structure, comprising SiN$_4$ and MnN$_4$ tetrahedra, is also 
in good quantitative agreement; the average experimental (DFT) Si-N and Mn-N bond distances are 1.758\,\AA\ (1.758\,\AA) 2.144\,\AA\ (2.143\,\AA), respectively and both exhibit distorted tetrahedra with bond distortion indices of the order 10$^{-3}$. This lowers the ideal tetrahedral $T_d$ symmetry to $C_1$.
The corner-connectivity of the distorted tetrahedra produces buckled metal-N-metal bonds angles away from that expected for ideal tetrahedra in the wurtzite structure ($\sim109.5\,^\circ$).  This distortion along with the Si and Mn chemical order creates three distinct Si-N-Mn angles nearly along the principal axes ($a$, $b$, $c$): 
104.6\,$^\circ$ (106.8\,$^\circ$), 
112.8\,$^\circ$ (114.9\,$^\circ$), and
106.3\,$^\circ$ (106.2\,$^\circ$).
The Mn-N-Mn bond angles that mediate the superexchange are 
93.59\,$^\circ$ (93.72\,$^\circ$), and the average Mn-Mn distance is 
3.110\,\AA\ 
(3.128\,\AA).

\begin{figure}[t]
    \centering
    \includegraphics[width=0.5\textwidth]{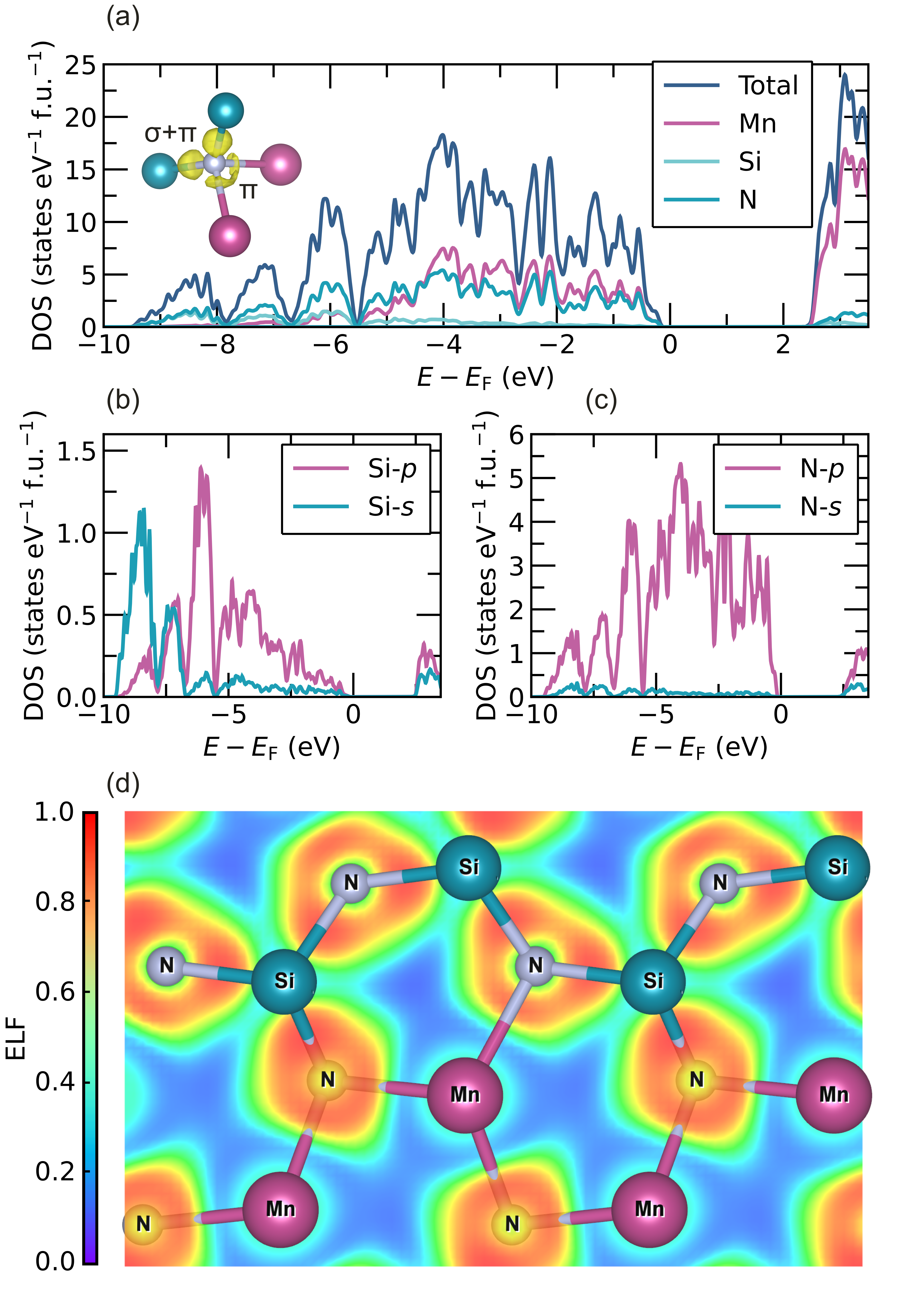}
    \caption{
    (a)  Noncollinear density of states (DOS) of  MnSiN$_2$ with the magnetic $Pc^\prime$ symmetry (\emph{vide infra}) with a band gap of  $E_\mathrm{gap}=2.33$~eV showing the overlap of Mn-$d$ and N-$p$ states. 
    (inset) 3D ELF (isocontour of 0.84) of the NMn$_2$Si$_2$ tetrahedron showing 
    the bonding localization basins comprise covalent $\sigma$ and $\pi$ Si-N bonding  and primarily Mn-N $\pi$ bonding.
    Band overlap of (b) Si-$p$ and Si-$s$ states, showing Si $sp^3$ hybridization, 
    and (c) N-$p$ and N$-s$ states throughout the valence band. 
    Two-dimensional ELF plot in the (100) plane showing variations in  covalent bonding across the SiN$_4$ and MnN$_4$ tetrahedral network.
    \label{fig2:estruct}
 }
\end{figure}

%
\autoref{fig2:estruct}a shows the relativistic electronic density of states obtained for  MnSiN$_2$ in its low-temperature magnetically ordered state, exhibiting $Pc^\prime$ symmetry, and discussed in detail in the next sections.
It is essentially a collinear G-type antiferromagnet (AFM-G), i.e., all adjacent Mn$^{2+}$ spins are anti-aligned to satisfy superexchange principles for the $d^5$ ions, with small spin canting. 
The charge state is insulating with a calculated band gap $E_{\rm gap}=2.33$~eV, consistent with the red color of the synthesized material. The calculated gap is reduced to approximately 0.5\,eV when no Hubbard correction is used.

The valence band (VB) spans approximately from -10\,eV to 0\,eV with a pseudogap near -5.5\,eV that separates the Mn-dominated states at the band edge from those of Si further below (\autoref{fig2:estruct}a). 
The N $2p$ states are found throughout and the deep N $2s$ states appear below -10\,eV (not shown).
The VB region comprises hybridized Si $3p$ and $3s$ states (\autoref{fig2:estruct}b) forming $sp^3$ orbitals overlapping with N $2p$ states (\autoref{fig2:estruct}c) to form single $\sigma$ bonds. 
The distribution of states is similar to the electronic structure reported for the thermodynamically stable $\beta$-Si$_3$N$_4$ nitride with tetrahedral bonding (average Si-N bond distance of 1.738\,\AA) \cite{PhysRevB.51.17379}, indicating strong covalent bonding in the SiN$_4$ tetrahedra.
Both $\sigma$- and $\pi$-bonding interactions along the Si-N bonds are visible in the 3D electron localization function (ELF) presented as an inset to \autoref{fig2:estruct}a, where ELF values greater than 0.7 are indicative of bonding regions. 
The covalent two-electron bond is highly localized between the atoms.
Rather than exhibiting a spherical shape \autoref{fig2:estruct}a (inset), the bond density forms a puckered dome from the $\pi$ back-bonding interaction through an orthogonal $p$ orbital that reduces the large effective nuclear charge of the nominal N$^{3-}$ anion.
These perturbations to the spherical $\sigma$ bond along the Si-N bond direction are further discernible in the two-dimensional ELF in the (100) plane (\autoref{fig2:estruct}d).

The low-energy portion of the valence band comprises primarily Mn $3d^5$ electrons (nominally $e^2t_2^3$ majority spins) hybridized with N $2p$ states from -6 eV to 0\,eV, while the minority spin states are unoccupied as described in a nonrelativistic description. 
%
%
The high-spin insulating state is robust to the details of our DFT simulations and the strong Mn-N interactions arise from the ligand character of the N$^{3-}$ anion.
While forming $dp\sigma$ antibonding states with the Mn $3d$ orbitals, we also find substantial $dp\pi$ interactions that suppress the number of states at the valence band edge.
We find additional covalent $\pi$ interactions in the 3D ELF along the Mn-N bond (\autoref{fig2:estruct}a, inset), where a torus of bonding electrons are found transverse to the bond direction.
This appears as increased localization in the 2D (100) projection as well in the form of a peanut-shape perpendicular to the Mn-N bond axis (\autoref{fig2:estruct}d).
The covalency reduces the magnetic moments for Mn, which we calculate as 4.37\,$\mu_\mathrm{B}$ per Mn atom in the collinear AFM-G configuration. 
As a consequence, the nitrogen-mediated superexchange should be enhanced in MnSiN$_2$ owing to nitrogen acting as both a $\sigma$-donor and $\pi$-donor. 
Although the O$^{2-}$ anion acts as a similar ligand, the N$^{3-}$ anion is a stronger $\pi$-donor owing to the larger size and lower electronegativity of nitrogen compared to oxygen.
These features allow better overlap with the metal orbitals, partial donation of lone pair of electrons into empty $\pi^*$ orbitals of the metal, and shorter metal-nitrogen bond lengths, which enhance the $\pi$ contributions to the superexchange.

\subsection{Magnetic properties}

\begin{figure}
	\centering
	\includegraphics[width=0.5\textwidth,trim=0 30 0 0, clip]{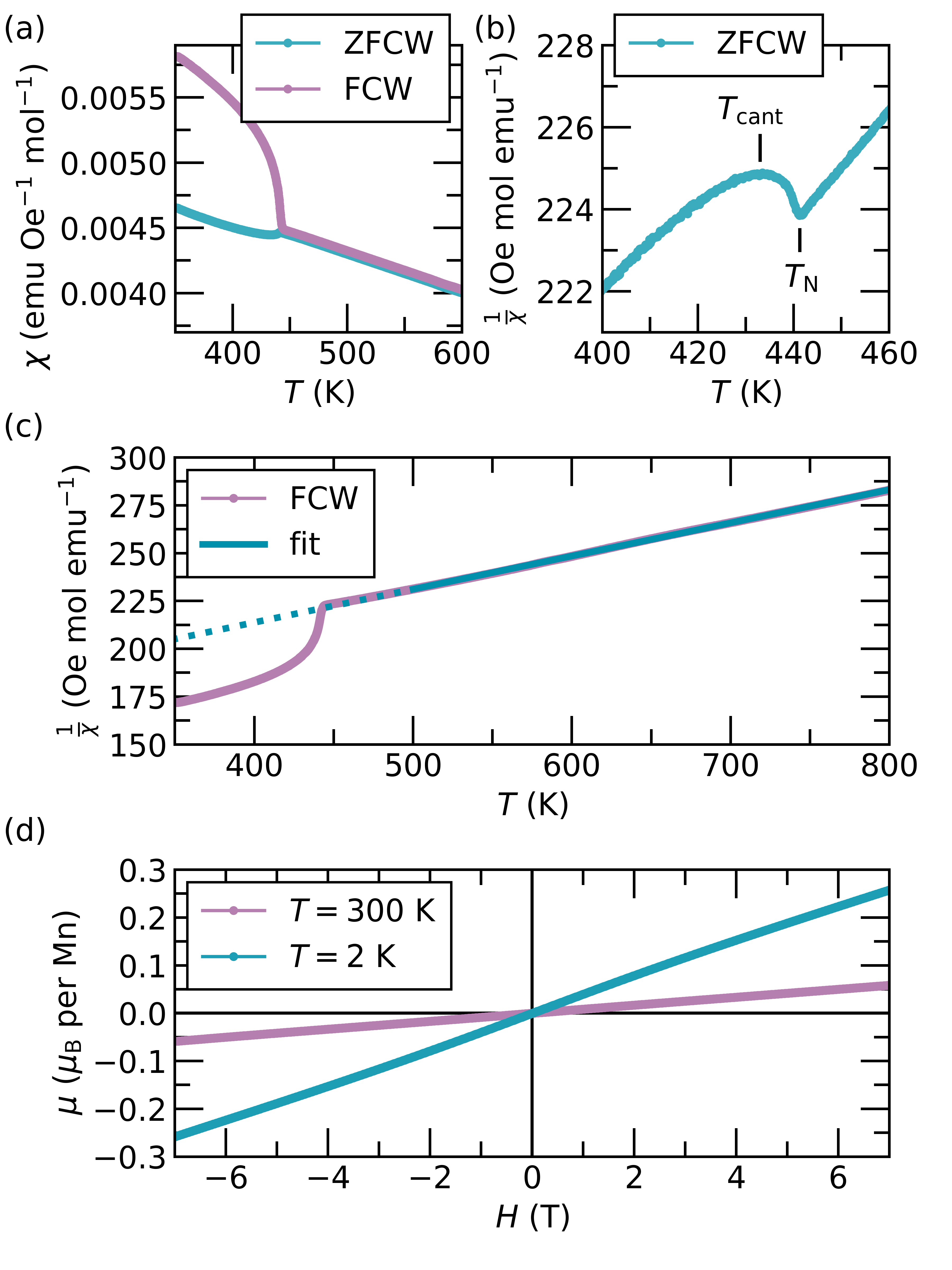}
	\caption{(a) High temperature susceptibility measured at an applied field of $H=0.1$~T showing the N\'{e}el ordering and canting temperatures. ZFCW indicates a zero field cooled measurement carried out upon warming. FCW indicates a field cooled measurement upon warming. (b) Enlarged inverse susceptibility highlighting $T_{\rm N}$ and $T_{\rm cant}$. (c) Curie-Weiss fit to high temperature susceptibility from FCW measurement. The $R^2=0.9998$ value indicates an excellent fit and a spin-only moment of $\mu_{\rm s}=6.79$~$\mu_{\rm B}$ and a Curie temperature of $\theta_{\rm CW}=-832$~K are found. (d) Magnetic moment $\mu$ as a function of the applied magnetic field $H$ at $T=300$~K and $T=2$~K.}
	\label{fig3:struc_mag}
\end{figure}

Temperature-dependent magnetic susceptibility measurements were performed and reveal a high Ne\'{e}l ordering temperature of $T_{\rm N}=443$~K (\autoref{fig3:struc_mag}a). 
An additional magnetic transition also appears at $T_{\rm cant}=433$~K (\autoref{fig3:struc_mag}b), which we attribute to the canting of the magnetic moments forming the AFM-G ordered state (\emph{vide infra}). 
Both critical temperatures are in good agreement with previously reported transition temperatures \cite{Esmaeilzadeh06}. 
A Curie-Weiss (CW) fit to the high-temperature magnetic susceptibility in the range of $T=500$~K to $T=800$~K result in $\theta_{\rm CW}=-832$~K and $\mu_{\rm s}=6.79$~$\mu_{\rm B}$ (\autoref{fig3:struc_mag}b). 
The negative value of $\theta_{\rm CW}$ indicates dominant antiferromagnetic interactions and weak magnetic frustration with $|\theta_{\rm CW}| \approx 2 T_{\rm N}$. 
The spin-only moment is larger than expected for a high-spin free Mn$^{2+}$ ion,  $\mu_{\rm s}=5.92\,\mu_{\rm B} =\sqrt{4S(S+1)}$ ($S=5/2$), but this is unlikely due to an orbital contribution as our DFT calculations show the orbital moment is nearly fully quenched as expected for a half-filled $d$ manifold in a (distorted) tetrahedral environment. The discrepancy is anomalous, and the local moment from our CW analysis is more consistent with a $J=3$ moment, which implies either unaccounted orbital effects or an invalid thermal regime of CW analysis owing to the large exchange field. Esmaeilzadeh et al.\cite{Esmaeilzadeh06} find a similarly large moment of $\mu_{\rm s}=6.87\,\mu_{\rm B}$ in their CW fit and attribute the discrepancy to an invalid thermal regime for the CW analysis. 

\autoref{fig3:struc_mag}c shows the magnetic moment $\mu$ as a function of the applied magnetic field $H$. The linear curves at $T=300$~K and $T=2$~K indicate the presence of canted antiferromagnetic order that is stable up to magnetic fields of $\mu_0 H=7$~T. It is possible that a small manganese oxide impurity, such as Mn$_3$O$_4$ or other dilute fraction of paramagnetic impurities lead to the slight s-shape of the curve at $T=2$~K. Additionally, a small coercive field of $H_{\rm c}=12$~mT at $T=2$~K suggests the presence of a magnetic impurity. This impurity is too small to be detected in synchrotron x-ray or neutron diffraction (\autoref{fig1:diff}).

\subsection{Magnetic structure}

\autoref{fig1:diff}c shows the powder neutron diffraction data collected at $T=1.5$\,K, which we acquired to fully determine the  magnetic spin structure of MnSiN$_2$. 
The software package SARAh was used to begin the initial refinement as it can determine the irreducible representation (irreps) and basis vectors for a commensurate magnetic structure consistent with measured magnetic reflections. 
The space group $P1$ was initially used for the magnetic phase and basis vectors from individual irreps $\Gamma_1$ to $\Gamma_4$ were tested.  The intensities of the magnetic reflections were poorly captured ($R_{\rm wp}>27$). 
To improve the magnetic structure refinement, a combination of basis vectors from multiple irreps was necessary. 
This constraint indicates that the symmetry of the magnetic phase is likely lower than that of the nuclear symmetry and led us to perform a series of noncollinear DFT calculations, allowing for multiple spin configurations, to guide the magnetic refinements.

\begin{figure}
    \centering
    \includegraphics[width=0.5\textwidth]{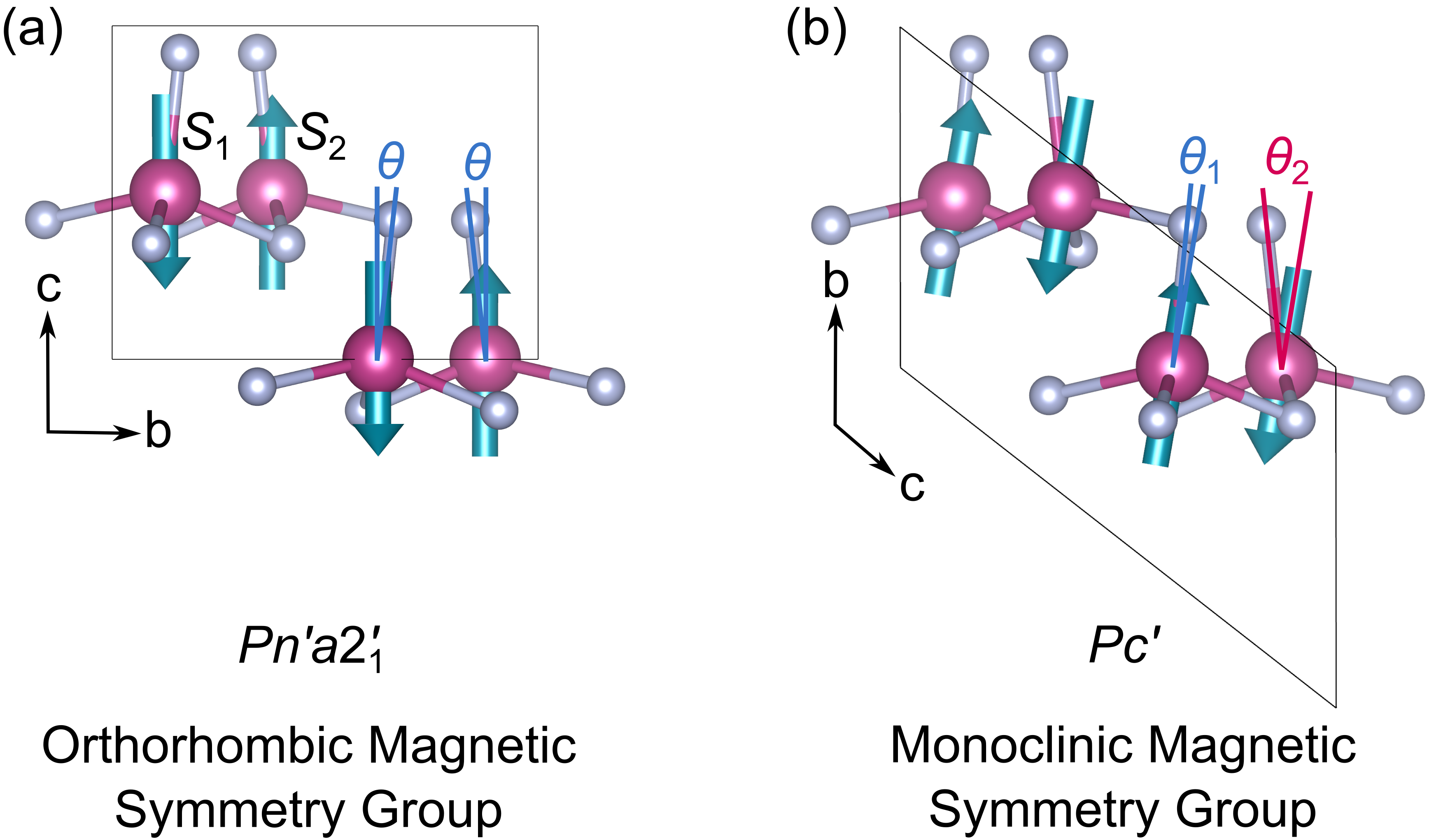}
    \caption{Spin orientations for Mn spins $S_1$ and $S_2$ in the 
    (a) orthorhombic magnetic group with $\theta=6.2\,^\circ$, and 
    (b) monoclinic magnetic group with $\theta_1=5.6\,^\circ$ and $\theta_2=12.2\,^\circ$, where 
    $\theta$ is defined as the angle between the Mn-N bond director and large AFM-G spin component from the noncollinear DFT calculations. 
    The local variation in $\theta$ from the orthorhombic to mononclinic structure reflects the rotation of the AFM component of the spins into the $bc$ plane; the canting angle present in the $Pc^\prime$ phase is given by $\theta-\theta_1=0.6\,^\circ$.
    }
    \label{fig4:maggroups}
\end{figure}

We first computationally explored various magnetic configurations and found that the AFM-G order is the most stable and strongly favored over other AFM configurations. The AFM-C order is 67\,meV per Mn higher in energy and the AFM-A order is 99\,meV per Mn higher in energy as shown in Ref.\ \onlinecite{magstructures}.
The ferromagnetic configuration is strongly disfavored (200\,meV per Mn higher in energy). 
We then surveyed the orthorhombic magnetic symmetries compatible with the crystallographic symmetry to determine the easy direction of the spins. 
Three possible magnetic groups compatible with $Pna2_1$ are: 
$Pn^\prime a^\prime 2_1$ (number 33.148), $Pna^\prime2_1^\prime$ (33.147), and $Pn^\prime a2_1^\prime$ (33.146), and can be found in Ref.\ \onlinecite{magstructures}.
Each of these magnetic groups permits a noncollinear antiferromagnetic alignment in the $ab$, $bc$, and $ac$ planes, respectively, with a ferromagnetic component orthogonal to the AFM direction. 
After relaxing various magnetic moment amplitudes compatible with each magnetic symmetry, we found a 17\,meV/Mn lower energy orthorhombic magnetic structure, 
much smaller than the energy differences between collinear spin arrangements, for the collinear AFM-G spin structure  with the spins along the $c$ axis (\autoref{fig4:maggroups}a), i.e., oriented along the vertex of the MnN$_4$ tetrahedra, and described by $Pn^\prime a2_1^\prime$. 
The two Mn spins, $S_1$ and $S_2$, are collinear and there is no spin canting; the angle $\theta$ formed between the spin direction and the Mn-N bond are identical for the two spins. 
In addition, the AFM spin component in the $a$ direction is essentially zero following relaxation; therefore, although the magnetic symmetry permits a weak FM component, the Dzyaloshinskii-Moriya interaction (DMI) proportional to $S_1\times S_2=0$; thus, there is no net FM along the $b$ axis.
The other orthorhombic magnetic groups were slightly higher in energy (of the order 70-85\,$\mu$eV/Mn).

We next lowered the magnetic symmetry further while maintaining the atomic coordinates of the orthorhombic crystal structure by starting with large AFM-G spins aligned along $bc$ and a FM component along $a$.
We then relaxed the spin structure and found two nearly degenerate monoclinic magnetic structures: 
$Pc$ (7.24)  with $9.5\,^\circ$ rotation of the spins relative to the $z$ direction along the $y$ axis, which is 18\,$\mu$eV/Mn and a minor FM spin component along the $x$ axis, and $P{c^\prime}$  (7.26) with a $10.0\,^\circ$ rotation  relative to the $z$ direction; however, with the spins along $y$ axis (20\,$\mu$eV/Mn higher in energy). 
(Here we use Cartesian directions to facilitate comparison of the monoclinic symmetries  with the orthorhombic magnetic groups, i.e., $z$ in the monoclinic structure is the same as the $c$ direction of the orthorhombic structure (\autoref{fig4:maggroups}.)
The DMI permits a weak-FM component along the $x$ direction of the order $5\times10^{-3}\,\mu_\mathrm{B}$ per Mn in the $Pc$ phase owing to low spin-orbit coupling strength in the system. 
In contrast, $P{c^\prime}$  is strictly zero owing to a permutation in the minor component of the AFM order of $S_1$ and $S_2$ in the Mn-Mn pairs (\autoref{fig4:maggroups}b), i.e., one pair contributes a w-FM component along $+x$ while the other gives a $-x$ component.
Both monoclinic spin structures remove symmetry operations formerly present in the orthorhombic groups. 
Although the Mn ions with the same sign of the $z$ spin projection are related by symmetry, adjacent Mn spins $S_1$ and $S_2$ can no longer be mapped onto one another. 
This inequivalence arises from canting of the two spins, such that the angles between the spin direction and the Mn-N bonds are no longer the same, i.e., $\theta_1\neq\theta_2$ (\autoref{fig4:maggroups}b). 
In addition, if $\theta_1=\theta_2\neq0$ then there remains no crystallographic symmetry linking the two spins. 
The magnetic symmetry is then monoclinic; however, if $\theta_1=\theta_2=0$ as in \autoref{fig4:maggroups}a and measured with respect to the orthorhombic $c$ axis ($b$ monoclinic), then the orthorhombic symmetries are compatible with $Pn^\prime a2_1^\prime$.
These magnetic structures are available from Ref.\ \onlinecite{magstructures}. 

\begin{figure}
	\centering
	\includegraphics[width=0.5\textwidth,trim=0 20 0 0]{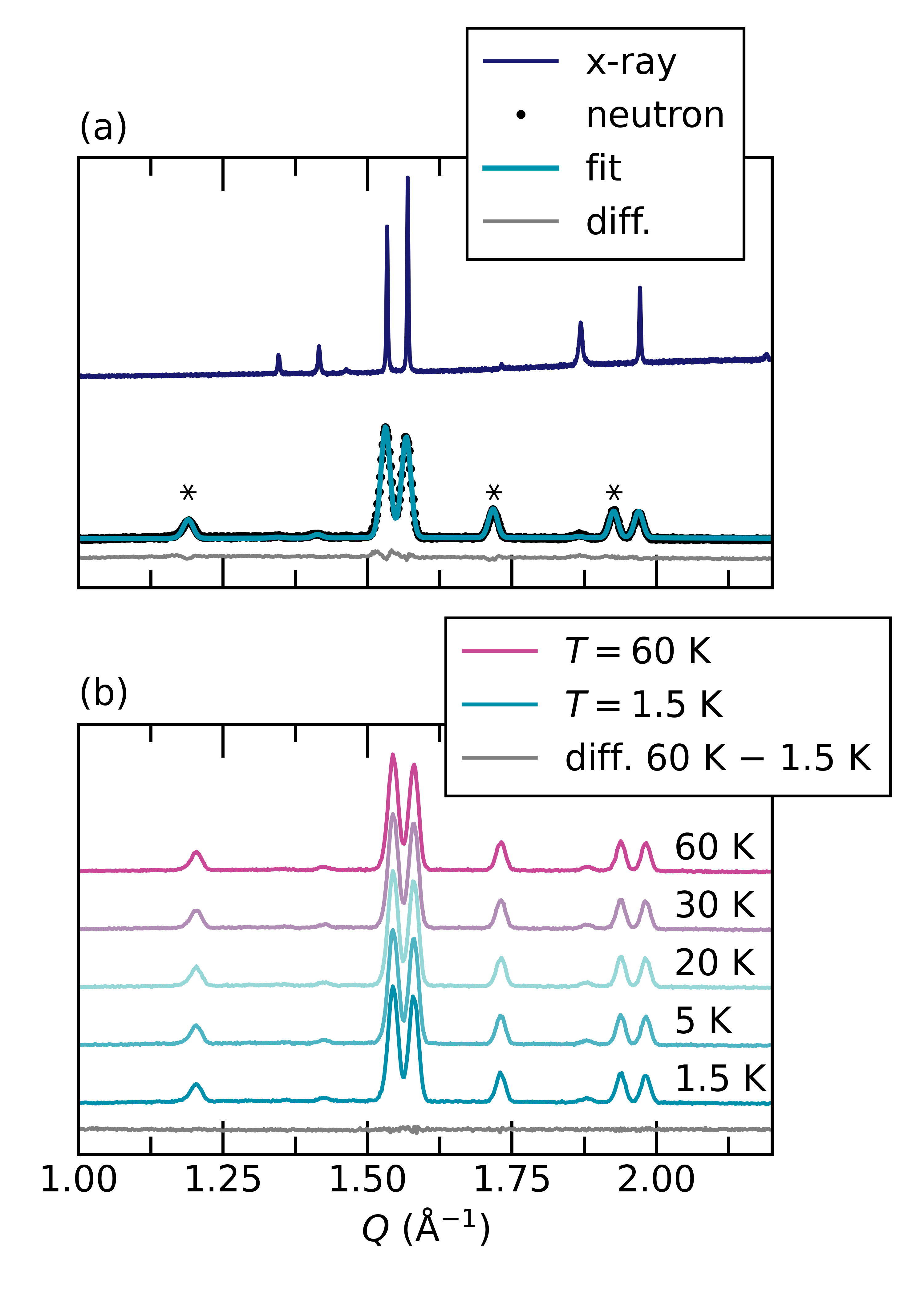}
	\caption{(a) Powder x-ray ($T=295$~K) and neutron diffraction ($T=1.5$~K) pattern at low $Q$ showing the additional magnetic Bragg peaks (marked by *) emerging from the magnetic phase. (b) Powder neutron diffraction patterns at low $Q$ and difference between data at $T=60$~K and data at $T=1.5$~K.}
	\label{fig5}
\end{figure}

We then performed a Rietveld fit of the powder neutron diffraction data at 1.5\,K using the DFT-calculated crystallographic structure of MnSiN$_2$ (space group $Pna2_1$) and the  $P_{c^\prime}$ magnetic symmetry with significant spin projections along the $z$ and $y$ axes \autoref{fig1:diff}c.
We found excellent agreement between our model and the measured 1.5\,K data without refining the magnetic vectors but allowing the lattice parameters and peak shapes to refine.
\autoref{fig5}a compares the low-$Q$ region of the neutron diffraction data to the x-ray diffraction data and clearly identifies three peaks (marked by *) as magnetic reflections. This procedure is necessary as the neutron diffraction experiment did not access the paramagnetic state due to the high ordering temperature of MnSiN$_2$.
We experimentally confirm that the ground state exhibits nominally $S=5/2$ spins moments aligned along the orthorhombic $c$ axis with a rotation angle of 10$^{\circ}\pm$1$^{\circ}$ away from the $c$-axis towards $b$.
The magnetic moment obtained from the neutron refinements in space group $Pc^\prime$ is $\mu_\mathrm{Mn}=4.37$~$\mu_\mathrm{B}$ (\autoref{tab:refin}). 
Although consistent with our DFT calculations, this value is smaller than expected for high-spin Mn$^{2+}$ ions and deduced from our CW fit. 
The measured moment is also larger than the value reported in Ref.\ \onlinecite{Esmaeilzadeh06} with $\mu_\mathrm{Mn}=3.55$~$\mu_\mathrm{B}$.
One possible reason for this smaller ordered moment value is the enhanced Mn-N bond covalency and a modified form factor for moment distribution.  A second possibility is the presence of enhanced fluctuation effects due to unaccounted frustration arising from longer-range interactions across the Mn sublattice. Future inelastic neutron scattering measurements on single crystals capable of characterizing these fluctuations and performing a total moment sum rule analysis are an appealing next step.

We further focus on the low-$Q$ region of the neutron powder diffraction data at multiple temperatures below the critical temperatures (\autoref{fig5}b). The difference between the data at $T=60$~K and $T=1.5$~K indicates that no additional continuous canting of the antiferromagnetic spin structure is observed when lowering the temperature to $T=1.5$~K. This, in combination with the absence of magnetic transitions below the canting temperature of $T_{\rm cant}=433$~K in susceptibility measurements, indicates that the low-temperature spin structure of MnSiN$_2$ is formed at $T_{\rm cant}=433$~K and remains unchanged to $T=1.5~K$. This is a remarkable result highlighting the strong magnetic superexchange resulting from the nitrogen-mediated superexchange coupling. 

\begin{figure}
    \centering
    \includegraphics[width=0.5\textwidth]{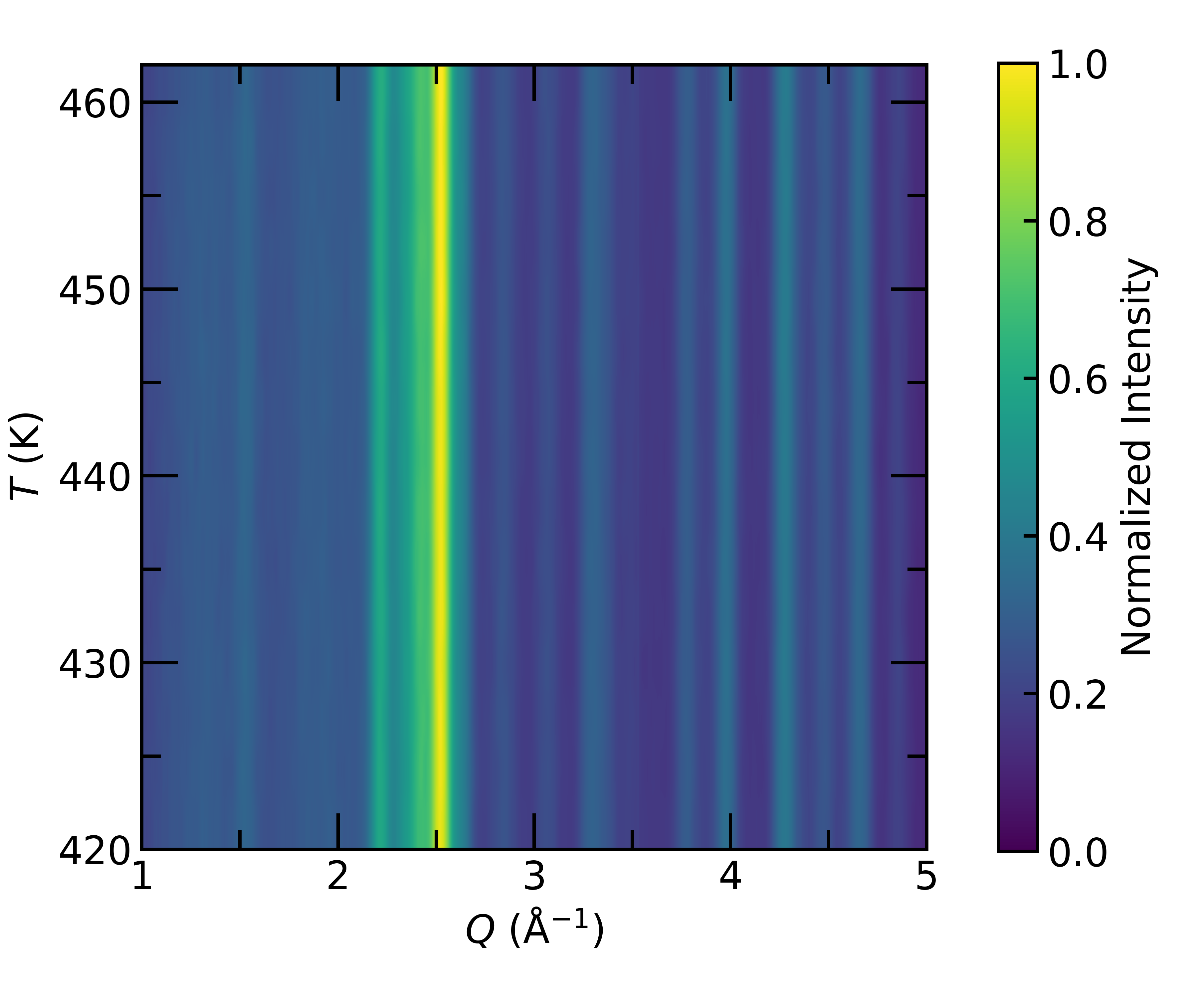}
    \caption{Temperature dependent powder x-ray diffraction spanning the magnetic ordering transition regime from $T=420$~K to $T=462$~K.}
    \label{fig_t_dep}
\end{figure}

To understand why MnSiN$_2$ orders as a collinear AFM then transforms into a canted AFM, we performed a computational experiment whereby we cooperatively modulated the nitride ligands reducing the amplitude of their distortion that buckles the Mn-N-Si network. 
Upon reducing the equilibrium bond angles obtained for the stable $Pc^\prime$ structure, i.e., reducing the buckling by 90\,\% to make the tetrahedra  more regular, we found that the $Pn^\prime a2_1^\prime$ and $Pc^\prime$ become energetically degenerate.
Thus, we hypothesize that the separation in $T_{\rm N}=443$ and $T_{\rm cant}$ could be due to temperature-dependent variations in the nitrogen Wyckoff site.
To that end, \autoref{fig_t_dep} shows our high-temperature PXRD across the magnetic transitions. 
Although the Mo X-ray source and geometry broaden the peaks, we find there is no crystallographic orthorhombic to monoclinic symmetry lowering.
There are slight changes to the intensities, and we fit the data to find very minimal changes in the atomic positions between 420\,K and 462 \,K.
There were more significant but still small displacements between 1.5 K and 420 K, and the Mn-N-Mn bond angle decreased from 97.5\,$^\circ$ to 96.7\,$^\circ$.
These changes are much smaller than predicted to be necessary by our simulation, suggesting weak magnetostructural coupling and unlikely that new exchange paths are activated. 

Spin canting in an antiferromagnet below its N\'eel temperature can also be caused by the presence of an external magnetic field, structural defects within the crystal lattice, or higher-order exchange interactions.
We have minimized contributions from these effects in our experiments and they are controlled or nonexistent in our simulations.
Therefore, we contend that  the transition from the collinear to canted AFM state is due to thermal fluctuations.
The energy scale between competing ordered states is comparable to the energy gain obtained from the spin canting. Thermal energy thus acts to reduce the macroscopically observable spin canting and it becomes more pronounced as dynamical spin fluctuations decay.

\section{Conclusion}

We have synthesized the polycrystalline material MnSiN$_2$ to reexamine its magnetic structure using powder neutron diffraction. Our magnetic measurements reveal a N\'{e}el ordering temperature of $T_{\rm N}=443$~K and an additional magnetic canting transition at $T_{\rm cant}=433$~K. Irreducible representation analysis showed that the magnetic reflections could not be well captured using basis vestors form one of the individual irreducible representations $\Gamma_1$ to $\Gamma_4$. The magnetic refinement required a combination of basis vectors from multiple representations, indicating  symmetry of the magnetic phase is lower than the nuclear symmetry. A series of noncollinear DFT calculations show a symmetry reduction to a canted antiferromagnetic G-type spin configuration with the spins rotated in the $bc$ plane is favored ($Pc^\prime$ symmetry). This model of the magnetic structure fits the magnetic Bragg peaks observed at $T=1.5$\,K well. We conclude that the ground state magnetic structure of MnSiN$_2$ is a G-type antiferromagnet with a $10\,^\circ$ rotation of the spins away from the crystallographic $c$ direction  (orthorhombic setting) and 0.6\,$^\circ$ canting.

Our density functional theory calculations support an experimental bandgap of $E_{\rm gap}=2.33$~eV, consistent with the red color of the material. The nitrogen $2p$ states in MnSiN$_2$ span a large portion of the valence band and overlap with silicon $sp^3$ hybridized orbitals forming strong $\sigma$ bonds. This creates SiN$_4$ tetrahedra with strong convalent bonding. The manganese $3d^5$ electrons hybridize with the nitrogen $2p$ states and strong Mn-N interactions arise from the $N^{3-}$ ligand acting as a strong $\pi$-donor. The increased covalent $\pi$ interactions along the Mn-N bond result in a slightly reduced magnetic moment of 4.37\,$\mu_{\rm B}$ per Mn atom, which is consistent with the neutron diffraction data and previous reports in literature. The nitrogen anion acting as $\sigma$- and $\pi$-donors enhances the strength of the nitrogen-mediated superexchange pathways and leads to the high N\'{e}el ordering temperature in MnSiN$_2$.

\begin{acknowledgments}
This work was supported by the Air Force Office of Scientific Research under award number FA9550-23-1-0042. 
The research reported here made use of the shared facilities of the Materials Research Science and Engineering Center (MRSEC) at UC Santa Barbara: NSF DMR–2308708.
The UC Santa Barbara MRSEC is a member of the Materials Research Facilities Network (www.mrfn.org).
Use of the Advanced Photon Source at Argonne National Laboratory was supported by the US Department of Energy, Office of Science, Office of Basic Energy Sciences, under Contract No. DE-AC02-06CH11357. This research used resources at the High Flux Isotope Reactor, a DOE Office of Science User Facility operated by the Oak Ridge National Laboratory. The computational work  was supported by the National Science Foundation (NSF) Grant no. DMR-2011208 and used resources of the National Energy Research Scientific Computing Center (NERSC), a U.S. DOE Office of Science User Facility located at Lawrence Berkeley National Laboratory, operated under Contract no.\ DE-AC02-05CH11231.
\end{acknowledgments}

\bibliography{references}
\end{document}